\documentclass[runningheads,a4paper,12pt]{llncs}
\usepackage[utf8]{inputenc}
\usepackage[T1]{fontenc}
\usepackage{amsmath, amsfonts}
\usepackage{amssymb}
\usepackage{geometry}
\usepackage{url}
\urldef{\mailsa}\path|{alfred.hofmann, ursula.barth, ingrid.haas, frank.holzwarth,|
\urldef{\mailsb}\path|anna.kramer, leonie.kunz, christine.reiss, nicole.sator,|
\urldef{\mailsc}\path|erika.siebert-cole, peter.strasser, lncs}@springer.com|

\usepackage{newtxtext}
\usepackage{newtxtt}
\usepackage{textgreek}
\usepackage{graphicx}
\usepackage{latexsym}
\usepackage[colorlinks=true]{hyperref}
\usepackage{xspace}
\usepackage{paralist}
\usepackage{caption,subcaption}
\usepackage{tikz}
\usepackage{calc}

\usetikzlibrary{arrows,decorations.pathmorphing,backgrounds,positioning,fit} 
\usetikzlibrary{shapes.geometric}

\usetikzlibrary{intersections}
\usepackage{tkz-euclide}
\usetkzobj{all}
\usetikzlibrary{calc}
\usepackage[ruled, linesnumbered,vlined]{algorithm2e}

\newcommand{\decimaxsum}{\textsc{DeciMaxSum}\xspace}

\newcommand\trigger{\ensuremath\Theta\xspace}%\mathtt{trigger}}
\newcommand\candidates{\ensuremath\Phi\xspace}%\mathtt{sel\_var}}
\newcommand\variables{\ensuremath\Upsilon\xspace}%\mathtt{sel\_var}}
\newcommand\values{\ensuremath\Lambda\xspace}%\mathtt{sel\_val}}
\newcommand\eqdefU{\ensuremath{\mathop{\overset{\text{def}}{=}}}}
\newcommand\eqdef{\mathop{\overset{\text{def}}{\resizebox{\widthof{\eqdefU}}{\heightof{=}}{=}}}}

\DeclareMathOperator*{\argmax}{argmax}

\begin{document}

\mainmatter  % start of an individual contribution

\title{Improving Max-Sum through Decimation to Solve Loopy Distributed Constraint Optimization Problems}
\titlerunning{\decimaxsum: Decimation in MaxSum}

\author{J. Cerquides\inst{1} \and R. Emonet\inst{2} \and G. Picard\inst{3} \and J.A. Rodriquez-Aguilar\inst{1}% Alfred Hofmann%
% \thanks{Please note that the LNCS Editorial assumes that all authors have used
% the western naming convention, with given names preceding surnames. This determines
% the structure of the names in the running heads and the author index.}%
% \and Ursula Barth\and Ingrid Haas\and Frank Holzwarth\and\\
% Anna Kramer\and Leonie Kunz\and Christine Rei\ss\and\\
% Nicole Sator\and Erika Siebert-Cole\and Peter Stra\ss er
}
\authorrunning{Cerquides \emph{et al.}}

\institute{IIIA-CSIC, Campus UAB, 08193 Cerdanyola, Catalonia, Spain \email{\{cerquide,jar\}@iiia.csic.es}
  \and 
  Université de Lyon, Laboratoire Hubert Curien UMR CNRS 5516, France\\ \email{remi.emonet@univ-st-etienne.fr}
  \and
  MINES Saint-Etienne, Laboratoire Hubert Curien UMR CNRS 5516, France\\ \email{picard@emse.fr}
}


%\toctitle{Lecture Notes in Computer Science}
%\tocauthor{Authors' Instructions}

%\author{%M. Legendre\institute{Univ Lyon, MINES Saint-Etienne, CNRS, Laboratoire Hubert Curien UMR 5516, F-42023 Saint Etienne, France, email: \{marc.legendre,picard\}@emse.fr} \and J. Cerquides\institute{IIIA-CSIC, Campus UAB, 08193 Cerdanyola, Catalonia, Spain, email: \{cerquide,jar\}@iiia.csic.es} \and R. Emonet\institute{Univ Lyon, UJM-Saint-Etienne, CNRS, Laboratoire Hubert Curien UMR 5516, F-42023 Saint Etienne, France, email: remi.emonet@univ-st-etienne.fr} \and G. Picard\up{1} \and J.A. Rodr\'i{}guez-Aguilar\up{2}
%}

\maketitle

\begin{abstract}
In the context of solving large distributed constraint optimization problems (DCOP), belief-propagation and approximate inference algorithms are candidates of choice. However, in general, when the factor graph is very loopy (i.e. cyclic), these solution methods suffer from bad performance, due to non-convergence and many exchanged messages. As to improve performances of the Max-Sum inference algorithm when solving loopy constraint optimization problems, we propose here to take inspiration from the belief-propagation-guided decimation used to solve sparse random graphs ($k$-satisfiability). We propose the novel \decimaxsum method, which is parameterized in terms of policies to decide when to trigger decimation, which variables to decimate, and which values to assign to decimated variables. Based on an empirical evaluation on a classical BP benchmark (the Ising model), some of these combinations of policies exhibit better performance than state-of-the-art competitors.
\end{abstract}

\section{Introduction}
\label{sec:introduction}

In the context of multi-agent systems, distributed constraint optimization problems (DCOP) are a convenient to model coordination issues agents have to face, like resource allocation, distributed planning or distributed configuration. In a DCOP, agents manage one or more variables they have to assign a value (e.g. a goal, a decision), while taking into account constraints with other agents. Solving a DCOP consists in making agents communicate as to minimize the violation of these constraints. Several solution methods exist to solve such problems, from complete and optimal solutions, to approximate ones. When dealing with larger scales (thousands of variables), approximate methods are solutions of choice. Indeed, complete methods, like ADOPT or DPOP, suffer exponential computation and/or communication cost in general settings \cite{p.j._modi_adopt:_2005,Petcu2005}. As a consequence, in some large settings, approximate methods are better candidates, as evidenced by the extensive literature on the subject (see \cite{CerquidesFMR14} for a complete review). One major difficulty for approximate method to solve DCOP is the presence of cycles in the constraint graph (or factor graph). Among the aforementioned methods, inference-based ones, like Max-Sum \cite{Farinelli:2008:DCL:1402298.1402313} and its extensions like \cite{Rogers2011730}, have demonstrated good performance even on loopy settings. However, there exists some cases, with numerous loops or large induced width of the constraint graph, where they perform badly, which translates into a larger number of messages, a longer time to convergence and a final solution with bad quality. 

One original approach to cope with loopy graphs is to break loops by \emph{decimating} variables during the solving process. Decimation is a method inspired by statistical physics, and applied in belief-propagation, which consists in fixing the value of a variable, using the marginal values as the decision criteria to select the variable to decimate \cite{mezard09}. The decimation is processed regularly after the convergence of a classical belief-propagation procedure. In \cite{DBLP:journals/corr/abs-0709-1667}, decimation has been used in the constraint satisfaction framework, for solving centralized $k$-satisfiability problems \cite{DBLP:journals/corr/abs-0709-1667}. Inspired by this concept, we propose a general framework for applying decimation in the DCOP setting. Other works proposed Max-Sum\_AD\_VP as to improve Max-Sum performance on loopy graphs \cite{DBLP:conf/aamas/ZivanP12}. The idea is to perform the inference mechanism through an overlay directed acyclic graph, to remove loops, and to alternating the direction of edges at a fixed frequency as to improve the sub-optimal solution found with the previous direction. One mechanism within one of these extensions, namely value propagation, can be viewed as a temporary decimation.

Against this background, the main goal of this paper is to propose a general framework for installing decimation in Max-Sum for solving DCOP. More precisely, we make the following contributions:

\begin{enumerate}
\item We propose a parametric solution method, namely \decimaxsum{}, to implement decimation in Max-Sum. It takes three fundamental parameters for decimation: 
\begin{inparaenum}[(i)]
  \item a policy stating when to trigger decimation, \label{item:trigger}
  \item a policy stating which variables to decimate, and \label{item:variable}
  \item a policy stating which value to assign to decimated variables.\label{item:value}
  \end{inparaenum}
The flexibility of \decimaxsum comes from the fact that any policy from (\ref{item:trigger}) can be combined with any policy from (\ref{item:variable}) and (\ref{item:value}).
\item We propose a library of decimation policies; some inspired by the state-of-the-art and some original ones. Many combinations of policies are possible, depending on the problem to solve.
\item We implement and evaluate some of these combinations of decimation policies on classical DCOP benchmarks (meeting scheduling and Ising models), against state-of-the-art methods like standard Max-Sum and Max-Sum\_AD\_VP.
\end{enumerate}

The rest of the paper is organized as follows. Section~\ref{sec:background} expounds some background on DCOP and expounds the decimation algorithm from which our algorithm \decimaxsum{} is inspired. Section~\ref{sec:decimation} defines the general framework of \decimaxsum{}, and several examples of decimation policies. Section~\ref{sec:experiments} presents results and analyses of experimenting \decimaxsum{}, with different combinations of decimation policies, against Max-Sum and Max-Sum\_AD\_VP. Finally, % Section~\ref{sec:discussion} discusses the proposed framework in light of the experimental analyses, and
Section~\ref{sec:conclusion} concludes this paper with some perspectives.

%%% Local Variables:
%%% mode: latex
%%% mode: flyspell
%%% TeX-master: "decimation17report"
%%% End:

\section{Background}
\label{sec:background}

This section expounds the DCOP framework and some related belief-propagation algorithms from the literature are discussed concerning the mechanisms to handle cycles in constraint graphs.

\subsection{Disributed Constraint Optimization Problems}

One way to model the coordination problem between smart objects is to formalize the problem as a distributed constraint optimization problem. 

\begin{definition}[DCOP]
  \label{def:dcop}
A discrete \emph{Distributed Constraint Optimization
  Problem} (or DCOP) is a tuple $\langle \mathcal{A}, \mathcal{X}, \mathcal{D}, \mathcal{C}, \mu \rangle$, where:
\begin{inparaitem}
\item[] $\mathcal{A} = \{a_1,\ldots,a_{|A|}\}$ is a set of agents; 
\item[] $\mathcal{X} = \{x_1,\ldots, x_N\}$ are variables owned by the agents;
\item[] $\mathcal{D} = \{\mathcal{D}_{x_1},\ldots,\mathcal{D}_{x_N}\}$ is a set of finite
  domains, such that variable $x_i$ takes values in $\mathcal{D}_{x_i} = \{v_1,\ldots, v_k\}$;
\item[] $\mathcal{C} = \{u_1,\ldots,u_M\}$ is a set of soft constraints, where each $u_i$  defines a utility $\in \mathbb{R} \cup \{-\infty\}$ for each combination of assignments to a
  subset of variables $\mathcal{X}_i \subseteq \mathcal{X}$ (a constraint is initially known only to the agents involved);
\item[] $\mu: \mathcal{X} \rightarrow \mathcal{A}$ is a function mapping variables to their associated agent.
\end{inparaitem}
A \emph{solution} to the DCOP is an assignment $\mathcal{X}^* = \{x_1^*, \ldots, x_N^*\}$ to all variables that maximizes the overall sum of costs\footnote{Note that the notion of cost can be replaced by the notion of cost $\in \mathbb{R} \cup \{+\infty\}$. In this case, solving a DCOP is a minimization problem of the overall sum of costs.
}:
\begin{equation}
\sum_{m=1}^M u_m(\mathcal{X}_m)\label{eq:obj}
\end{equation}

\end{definition}
As highlighted in \cite{CerquidesFMR14}, DCOPs have been widely studied and applied in many reference domains, and have many interesting features:
\begin{inparaenum}[(i)]
\item strong focus on decentralized approaches where agents negotiate a joint solution through local message exchange;
\item solution techniques exploit the structure of the domain (by encoding this into constraints) to tackle hard computational problems;
\item there is a wide choice of solutions for DCOPs ranging from complete algorithms to suboptimal algorithms.
\end{inparaenum}

A binary DCOP can be represented as a constraint graph, where vertices represent variables, and edge represent binary constraints% , as pictured in Figure~\ref{fig:cg}
. In the case of n-ary constraints, a DCOP can be represented as a factor graph: an undirected bipartite graph in which vertices represent variables and constraints (called factors), and an edge exists between a variable and a constraint if the variable is in the scope of the constraint% , as pictured in Figure~\ref{fig:fg}
.

\begin{definition}[Factor Graph]
  A factor graph of a DCOP as in Def.~\ref{def:dcop}, is a bipartite graph $FG=\langle\mathcal{X},\mathcal{C},E\rangle$, where the set of variable vertices corresponds to the set of variables $\mathcal{X}$, the set of factor vertices corresponds to the set constraints $\mathcal{C}$, and the set of edges is $E = \{e_{ij}\ |\ x_i \in \mathcal{X}_j\}$.
\end{definition}

When the graph representing the DCOP contains at least a cycle, we call it a \emph{cyclic} DCOP; otherwise, it is \emph{acyclic}.

A large literature exists on algorithms for solving DCOPs which fall into two categories. On the one hand, \emph{complete algorithms} like ADOPT and its extensions \cite{modi05adopt}, or inference algorithms like DPOP \cite{Petcu2005} or ActionGDL \cite{Vinyals2010}, are optimal, but mainly suffer from expensive memory (e.g. exponential for DPOP) or communication (e.g. exponential for ADOPT) load --which we may not be able to afford in a constrained infrastructure, like in sensor networks. On the other hand, \emph{approximate algorithms} like Max-Sum \cite{Farinelli:2008:DCL:1402298.1402313} or MGM \cite{mgm2004} have the great advantage of being fast with a limited memory print and communication load, but losing optimality in some settings --e.g. Max-Sum is optimal on acyclic DCOPs, and may achieve good quality guarantee on some settings.

The aforementioned algorithms mainly exploit the fact that an agent's utility (or constraint's cost) depends only on a subset of other agents' decision variables, and that the global utility function (or cost function) is a sum of each agent's utility (constraint's cost). %In the following, we will discuss how cyclicity of constraint graphs (or factor graphs) impact the execution of these solution methods.
In this paper, we are especially interested in belief-propagation-based algorithms, like Max-Sum, where the notion of marginal values describes the dependency of the global utility function on variables. 
% \subsection{Impact of Cyclicity on DCOP Solving [0.5 p]}

% \subsubsection{Impact on Search Algorithms}

% \begin{itemize}
% \item No impact but exponential number of messages
% \end{itemize}

% \subsubsection{Impact on Inference Algorithms}

% \begin{itemize}
% \item Impact on message size (DPOP)
% \item Impact on number of messages + optimality (Max-Sum)
% \end{itemize}

\subsection{From Belief-Propagation to Max-Sum}

%As we propose to extend classical Max-Sum using decimation mechanisms, this section briefly presents belief-propagation and its DCOP derivation Max-Sum, before providing a summary of decimation-related methods in the next section.

Belief propagation (BP), i.e. sum-product message passing method, is a potentially distributed algorithm for performing inference on graphical models, and can operate on factor graphs representing a product of $M$ factors \cite{Mackay_2003_information}:
% \begin{equation}
%   \label{eq:1}
$F(x) = \prod_{m=1}^M f_m(\mathcal{X}_m)$
% \end{equation}
% For instance, factor graph in Figure~\ref{fig:mp} represents the factor $f_1(x_1)f_2(x_1,x_2,x_3)f_3(x_3)$
. The sum-product algorithm provides an efficient local message passing procedure to compute the marginal functions of all variables simultaneously. The marginal function, $z_n(x_n)$ describes the total dependency of the global function $F(x)$ on variable $x_n$: 
% \begin{equation}
%   \label{eq:2}
$  z_n(x_n) = \sum_{\{x'\},n'\neq n}F(\mathcal{X}_{n'})$.
BP operates iteratively propagating messages $m_{i\to j}$ (tables associating marginals to each value of variables) along the edges of the factor graph.% , as illustrated in Figure~\ref{fig:mp}. These messages depends on the type of the emitter:
% \begin{compactitem}
% \item messages \emph{from functions to variables}:
%   \begin{equation}
%     q_{n\to m}(x_n) = \displaystyle\prod_{m'\in\mathcal{V}(n)\setminus m}r_{m'\to n}(x_n)\label{eq:3}
%   \end{equation} where $\mathcal{V}$ is the set of indexes of variables connected to the function $f_m$, and
% \item messages \emph{from variables to functions}:
%   \begin{equation}
%     \label{eq:4}
%     r_{m\to n}(x_n) = \sum_{\mathcal{X}_m\setminus n}\left(f_m(\mathcal{X}_m)\prod_{n'\in\mathcal{F}(m)\setminus n}q_{n'\to m}(x_{n'})\right)
%   \end{equation}
% where $\mathcal{F}(n)$ is the set of indexes of functions connected to the variable $x_n$, and $\mathcal{X}_m\setminus n \myeq \{x_{n′} : n′ \in \mathcal{V}(m) \setminus n\}$.
% \end{compactitem}
 When the factor graph is a tree, BP algorithm computes the exact marginals and converge in a finite number a steps depending on the diameter of the graph \cite{Mackay_2003_information}. Max-product is an alternative version of sum-product which computes the maximum value instead of the sum.%  in Equation~\ref{eq:4}:
% \begin{equation}
%     \label{eq:5}
%     r_{m\to n}(x_n) = \max_{\mathcal{X}_m\setminus n}\left(f_m(\mathcal{X}_m)\prod_{n'\in\mathcal{F}(m)\setminus n}q_{n'\to m}(x_{n'})\right)
%   \end{equation} and use a scaler $\alpha_{nm}$, such that $\sum_{x_n}q_{n\to m}(x_n) =1$ to cope with cyclic graphs in Equation~\ref{eq:3}:
% \begin{equation}
%     q_{n\to m}(x_n) = \alpha_{nm}\displaystyle\prod_{m'\in\mathcal{V}(n)\setminus m}r_{m'\to n}(x_n)\label{eq:6}
%   \end{equation}
% Thus, max-product computes $z_n(x_n)$ differently than Equation~\ref{eq:2}:
% \begin{equation}
%   \label{eq:7}
%   z_n(x_n) = \displaystyle\max_{\mathcal{X}_m\setminus n}\prod_{m=1}^Mf_m(\mathcal{X}_{m})
% \end{equation}
% and the marginal value for variable $x_n$ is $x_n^* = \operatorname{arg\,max}_{x_n}z_n(x_n)$.

Built as a derivative of max-product, Max-Sum is an approximate algorithm to solve DCOP \cite{Farinelli:2008:DCL:1402298.1402313}. The main evolution is the way messages are assessed, to pass from product to sum operator through logarithmic translation. % In Max-Sum, message from variable to factor are assessed as follows:
%  \begin{equation}
%     Q_{n\to m}(x_n) = \alpha_{nm} + \displaystyle\sum_{m'\in\mathcal{V}(n)\setminus m}R_{m'\to n}(x_n)\label{eq:8}
%   \end{equation}
% and messages for factor to variable are defined by:
%  \begin{equation}
%     \label{eq:9}
%     R_{m\to n}(x_n) = \max_{\mathcal{X}_m\setminus n}\left(u_m(\mathcal{X}_m)\sum_{n'\in\mathcal{F}(m)\setminus n}Q_{n'\to m}(x_{n'})\right)
%   \end{equation}
And as a consequence, Max-Sum computes an assignment $\mathcal{X}^*$ that maximizes the DCOP objective in Equation~\ref{eq:obj}. Depending on the DCOP to solve, Max-Sum may be used with two different termination rules:
\begin{inparaenum}[(i)]
\item continue until convergence (no more exchanged messages, because when a variables or a factor receives twice the same message from the same emitter it does not propagates);
\item propagate message for a fixed number of iterations per agent.
\end{inparaenum}
Max-Sum is optimal on tree-shaped factor graphs, and still perform well on cyclic settings. But there exist problems for which Max-Sum does not converge or converge to a sub-optimal state. In fact, on cyclic settings \cite
{Farinelli:2008:DCL:1402298.1402313} identify the following behaviors:
\begin{inparaenum}[(i)]
\item agents converge to fixed states that represent either the optimal solution, or a solution close to the optimal, and the propagation of messages ceases;
\item agents converge as above, but the messages continue to change slightly at each update, and thus continue to be propagated around the network;
\item neither the agents' preferred states, nor the messages converge and both display cyclic behavior.
\end{inparaenum}

As to improve Max-Sum performance on cyclic graphs, \cite{DBLP:conf/aamas/ZivanP12} proposed two extensions to Max-Sum:
\begin{inparaenum}[(i)]
\item Max-Sum\_AD which operates Max-Sum on a directed acyclic graph built from the factor graph, and alternates direction at a fixed rate (a parameter of the algorithm);
\item Max-Sum\_AD\_VP which operates Max-Sum\_AD and propagates current values of variables when sending Max-Sum messages so that factors receiving the value only consider this value instead of the whole domain of the variable.
\end{inparaenum}
These two extensions, especially the second one, greatly improves the quality of the solution: Max-Sum\_AD\_VP found solutions that approximate the optimal solution by a factor of roughly $1.1$ on average. However, the study does not consider the number of exchanged messages, or the time required to converge and terminate Max-Sum\_AD\_VP.

% \item Summary of Max-Sum (algorithm) \cite{Farinelli:2008:DCL:1402298.1402313}

% \item Summary of message assessment
% \item Max-sum\_AD\_VP \cite{DBLP:conf/aamas/ZivanP12}
% \end{itemize}

\subsection{BP-guided Decimation}

In this paper, we propose to take inspiration from work done in computational physics \cite{mezard09}, as to cope with cyclicity in DCOP. Notably, \cite{2007PNAS..10410318K} introduced the notion of decimation in constraint satisfaction, especially $k$-satisfiability, where variables are binary, $x_i \in \{0, 1\}$, and each constraint requires $k$ of the variables to be different from a specific $k$-uple. Authors proposed a class of algorithms, namely \emph{message passing-guided decimation procedure}, which consists in iterating the following steps:
\begin{inparaenum}[(1)]
\item run a message passing algorithm, like BP ;
\item use the result to choose a variable index $i$ , and a
value $x_{i}^{*}$ for the corresponding variable;
\item replace the constraint satisfaction problem with the
one obtained by fixing $x_i$ to $x_i^*$.
\end{inparaenum}
The BP-guided decimation procedure is shown in Algorithm~\ref{alg:bp-guided}, whose performances are analysed in \cite{DBLP:journals/corr/abs-0709-1667,mezard09}.

\DontPrintSemicolon
\begin{algorithm}[tb]
  \KwData{A factor graph representing a $k$-satisfiability problem}
  \KwResult{A feasible assignment $\mathcal{X^*}$ or \texttt{FAIL}}
  \BlankLine
  initialize BP messages\;
  $\mathcal{U} \gets \emptyset$\;
  \For{$t = 1,\ldots,n$}{
    run BP until the stopping criterion is met\;\label{line:stop}
    choose $x_i \in \mathcal{X} \setminus \mathcal{U}$ uniformly at random\;\label{line:random}
    compute the BP marginal $z_i(x_i)$\;
    choose $x_i^*$ distributed according to $z_i$\;\label{line:sampling}
    fix $x_i = x_i^*$\;
    $\mathcal{U} \gets \mathcal{U} \cup \{x_i\}$\;
    simplify the factor graph\;
    if a contradiction is found, return \texttt{FAIL}\;
  }
  return $\mathcal{X}^*$
  \caption{The BP-guided decimation algorithm from \cite{DBLP:journals/corr/abs-0709-1667}\label{alg:bp-guided}}
\end{algorithm}

BP-guided decimation operates on the factor graph representing the $k$-satisfiability problem to solve. At each step, the variable to decimate is randomly chosen among the remaining variables. The chosen variable $x_i$ is assigned a value determined by random sampling according to its marginal $z_i$. After decimation, the factor graph is simplified: some edges are no more relevant, and factors can be sliced (columns corresponding to removed variables are deleted). In some settings, BP-guided decimation may fail, if random choices assign a value to a variable which is not consistent with other decimated variables.

Some comments can be made on this approach. First, \emph{relying on marginal values} is a key feature, and is the core of the ``BP-guided'' nature of this method. Marginal values are exploited to prune the factor graph. Second, while in the seminal work of \cite{DBLP:journals/corr/abs-0709-1667}, this procedure is used to solve satisfiability problems, the approach can easily be implemented to \emph{cope with optimization problems}. For instance, the inference library libDAI proposes an implementation of decimation for discrete approximate inference in graphical models \cite{Mooij_libDAI_10}, which was amongst the three winners of the UAI 2010 Approximate Inference Challenge\footnote{\url{http://www.cs.huji.ac.il/project/UAI10/}}. 

\subsection{State of a Factor Graph Representation}

The previous BP-based algorithm operates on factor graph representing the problem. ``Operates'' means that the algorithms create a data structure representing the factor graph which evolves with time : marginal values change, variables disappear, messages are sent/received, etc. Commonly, the logical representation of a factor graph is a set of nodes connected depending on the connectivity of the graph. Each such node has a state which stores some useful values.% , like the past received messages, the current marginal $z_i$, etc.

\begin{definition}
  The current state $FG^t$ % = \langle\state(\mathcal{X}),\state(\mathcal{C}),\state(E)\rangle$
  at time $t$ of a factor graph $FG=\langle\mathcal{X},\mathcal{C},E\rangle$ is the composition of all the current states of the data structures used by the BP-based algorithm to operate on the related factor graph, including the marginal values $z_i$, the messages $m_{i\to j}$, the set of decimated variables $\mathcal{U}$, and other algorithm-specific data. 
\end{definition}

We can consider that for a given problem, many factor graph states may exist. We denote $\mathfrak{S}$ the set of possible factor graph states, and $\mathfrak{S}(FG) \subset \mathfrak{S}$ the set of possible states for the factor graph $FG$.

%%% Local Variables:
%%% mode: latex
%%% mode: flyspell
%%% TeX-master: "decimation17report"
%%% End:

\section{\decimaxsum: Extending Max-Sum with Decimation}
\label{sec:decimation}

While mainly designed as a centralized algorithm and studied on $k$-SAT problems, BP-guided decimation could be utilized for solving DCOP with a few modifications. To the best of our knowledge, this approach has never been proposed for improving Max-Sum algorithm. Here we expound the core contribution of this paper, namely the \decimaxsum framework and its components.

\subsection{Principles}

The main idea is to extend the BP-guided decimation algorithm from \cite{DBLP:journals/corr/abs-0709-1667} in order to define a more general framework, in which other BP-based existing algorithms could fit.
First, the main focus is \emph{decimation}, which means assigning a value to a variable as to remove it from the problem. As the name suggests, there is no way back when a variable has been decimated --unlike search algorithms, where variable assignments can be revised following a backtrack, for instance. Therefore, triggering decimation is an impacting decision. This is why our framework is mainly based on answering three questions: \begin{inparaenum}[(i)]\item when is decimation triggered, \item which variable(s) to decimate, 
\item which value to assign to the decimated variable(s)?
\end{inparaenum}
Several criteria can be defined for answering each question, and the \textsc{DeciMaxSum} specifies such criteria as \emph{decimation policies}, that are fundamental parameters of the decimation procedure.

\begin{definition}[Decimation Policy]
  A \emph{decimation policy} is a tuple $\pi = \langle\trigger,\candidates,
  % \dependencies,
  \variables,\values\rangle$ where:
  \begin{itemize}
  \item $\trigger : \mathfrak{S} \to \{0,1\}$ is the condition to trigger the decimation process, namely the \emph{trigger policy},
  \item $\candidates : \mathfrak{S} \to 2^\mathcal{X}$ is a \emph{filter policy} which selects some candidate variables to decimate,
  %\item $\dependencies : \mathcal{X} \times \mathfrak{S} \to 2^\mathcal{X}$ defines the set of variables on which the decimation decision for a given variable depends on,
  \item $\variables : \mathcal{X} % \times 2^\mathcal{X}
    \times \mathfrak{S} \to \{0,1\}$ is the condition to perform decimation on a variable, namely \emph{perform policy},
  \item $\values : \mathcal{X} \times \mathfrak{S} \to \mathcal{D}_\mathcal{X}$ is the \emph{assignment policy}, which assigns a value to a given variable.
  \end{itemize}

\end{definition}

A rich population of decimation-based algorithm can be modeled through this framework by combining decimation policies. For instance, one can consider a \textsc{DeciMaxSum} instance, which
\begin{inparaenum}[(i)]\item triggers decimation once BP has converged, 
\item chooses randomly a variable to decimated within the whole set of non-decimated variables, and
\item samples the value of the decimated variable depending on its marginal values (used as probability distribution).
\end{inparaenum} By doing so, we result in the classical BP-guided decimation algorithm from \cite{DBLP:journals/corr/abs-0709-1667} . However, as many more decimation policies can be defined and combined, we fall into a more general framework generating a whole family of algorithms.

\subsection{\decimaxsum as an Algorithm}

We can summarize the \decimaxsum framework using Algorithm~\ref{alg:decimaxsum}. It is a reformulation of BP-guided decimation, parameterized with a decimation policy. Here decimation is not necessarily triggered at the convergence (or time limit) of BP. Criterion $\trigger$ may relies on other components of the state of the factor graph. Contrary to classical BP-guided decimation, there may be several variables to decimate at the same time (like in some variants of DSA or MGM) and that variables can be chosen in an informed manner (and not randomly), using criterion $\variables$. Values assigned to decimated variables, are not necessarily chosen stochastically, but are assigned using the function $\values$ that can be deterministic (still depending on the current state of the FG). Since, here we're not in the $k$-satisfiability case, but in an optimization case, there is no failure (only suboptimality), contrary to Algorithm~\ref{alg:bp-guided}. Finally, once all variables have been decimated, the output consists in decoding the state $FG^t$, i.e. getting the values assigned to decimated variables. This means that finally \decimaxsum is performing decoding while solving the problem, which is not a common feature in other DCOP algorithms, like classical Max-Sum or DSA. Indeed, once these algorithms halt, a decoding phase must be performed to extract the solution from the variables' states. 

\DontPrintSemicolon
\begin{algorithm}[tb]
  \SetCommentSty{textsl}
  \KwData{A factor graph $FG = \langle \mathcal{X},\mathcal{C}, E\rangle$, a decimation policy $\pi = \langle\trigger,\candidates,
  % \dependencies,
  \variables,\values\rangle$}
  \KwResult{A feasible assignment $\mathcal{X^*}$}
  \BlankLine
  initialize BP messages\;
  $\mathcal{U} \gets \emptyset$\;
  \While{$\mathcal{U} \neq \mathcal{X}$}{
    run BP until decimation triggers, i.e. $\trigger(FG^t) = 1$ \tcp*[r]{Sect.~\ref{sec:trigger}}
    choose variables to decimate, $\mathcal{X}' = \{x_i \in \candidates(FG^t) \ |\ \variables(x_i,% \dependencies(x_i,FG^t),
    FG^t)\} $\tcp*[r]{Sect.~\ref{sec:variables}}
    \For(\tcp*[f]{Sect.~\ref{sec:values}}){$x_i\in\mathcal{X}'$}{
      $x_i \leftarrow \values(x_i,FG^t)$\;
      $\mathcal{U} \gets \mathcal{U} \cup \{x_i\}$\;
      simplify $FG^t$ \tcp*[r]{remove variables, slice factors}
    }
  }
  return $\mathcal{X}^*$ by decoding $\mathcal{U}$
  \caption{The \decimaxsum framework as an algorithm\label{alg:decimaxsum}}
\end{algorithm}

While presented as a classical algorithm, let us note that decimation is meant to be implemented in a distributed and concurrent manner, depending on the decimation policy components. The rest of the section details and illustrates each of these decimation policies component with some examples.

\subsection{Triggering Decimation (\trigger\ criterion)}
\label{sec:trigger}

In the original approach proposed by \cite{DBLP:journals/corr/abs-0709-1667}, decimation is triggered once BP has converged. In a distributed settings and diffusing algorithms like BP, this can be implemented using termination detection techniques.
\begin{equation}
  \label{eq:converge}
  % \forall s \in \mathfrak{S}(FG), 
  \trigger_{\mathtt{converge}}(s) \eqdef \left\{
    \begin{array}{ll}
      1, & \text{if $s$ is \emph{quiescent}}\\
      0, & \text{otherwise}
    \end{array}\right.
\end{equation}

This trigger consists in detecting the \emph{quiescence} of the current state of the factor graph. This means no process is enabled to perform any locally controlled action and there are no messages in the channels \cite{lynch96}. Algorithms like \emph{DijkstraScholten} can detects such global state by implementing a send/receive network algorithm, based on the same graph than $FG$ \cite{lynch96}. Note that such techniques generates extra communication load for termination detection-dedicated messages.
% \end{example}

Due to the Max-Sum behavior on loopy factor graphs, convergence may not be reached \cite{DBLP:conf/aamas/ZivanP12}. The common workaround is to run BP for a fixed number of iterations in case there is no convergence. Setting this time limit (namely \texttt{LIMIT}) might be really problem-dependent.
\begin{equation}
  % \forall s \in \mathfrak{S}(FG),
  \trigger_{\mathtt{time}}(s) \eqdef \left\{
    \begin{array}{ll}
      1, & \text{if $time(s)$ = \texttt{LIMIT}}\\
      0, & \text{otherwise}
    \end{array}\right.
\end{equation}

In synchronous settings (all variables and factors are executed synchronously, step by step), getting the iteration number of the current state of the FG, $time(s)$, can done in a distributed manner, as usually done in Max-Sum. In the asynchronous case, one can either
\begin{inparaenum}[(i)]
\item use a shared clock, or 
\item count locally outcoming messages within each variables, and once a variable has sent a limit number of messages, decimation is triggered.
\end{inparaenum}

In some settings with strong time or computation constraints (e.g. sensor networks \cite{Farinelli:2008:DCL:1402298.1402313}, internet-of-things \cite{rust16ijcai}), waiting convergence is not affordable. Indeed, BP may generate a lot of messages. Therefore, we may consider decimating before convergence at a fixed rate (e.g. each $10$ iterations), or by sharing a fixed iteration budget amongst the variables (e.g. each $1000$ iterations divided by the number of variables). We can even consider a varying decimation speed (e.g. faster at the beginning, and lower at the end, as observed in neural circuits in the brain \cite{10.1371/journal.pcbi.1004347}).

\begin{equation}
  \label{eq:frequency}
% \forall s \in \mathfrak{S}(FG), 
\trigger_{\mathtt{frequency}}(s) \eqdef \left\{
    \begin{array}{ll}
      1, & \text{if } time(s) \bmod{f(s)} = 0\\
      0, & \text{otherwise}
    \end{array}\right.
\end{equation}
where $f$ is a function of the current state of the system, for instance :
\begin{itemize}
  \item $f(s) = \mathtt{RATE}$, with a predefined decimation frequency,
  \item $f(s) = \mathtt{BUDGET} / |\mathcal{X}|$, with a predefined computation budget,
  \item $f(s) = 2 \times time(s)$, for an decreasing decimation frequency.
\end{itemize}

Finally, another approach could be to trigger decimation once a loop in the FG is detected. Indeed, decimation is used here to cope with loops, so decimating variables, which could potentially break loops, seems a good approach.

\begin{equation}
  \label{eq:loop}
  \trigger_{\mathtt{loop}}(s) \eqdef \left\{
    \begin{array}{ll}
      1, & \text{if } \exists x_i\in\mathcal{X}, |loop(x_i)| > 1\\
      0, & \text{otherwise}
    \end{array}\right.
\end{equation}
where $loop(x_i)$ is the set of agents in the same first loop that $x_i$ just discovered. Detecting loops in the FG can be implemented during BP, by adding some metadata on the BP messages, like done in the DFS-tree construction phase of algorithms like DPOP or ADOPT.

% \begin{itemize}
% \item When to trigger decimation?
%   \begin{itemize}
%   \item After convergence (like
%     \cite{DBLP:journals/corr/abs-0709-1667}) [X]
%   \item At fixed rate [X]
%   \item With fixed budget divided among the variables (regular interval or not) [X]
%   \item When an agent detects a loop? [ ]
%   \end{itemize}
% \end{itemize}

\subsection{Deciding the Subset of Variables to Decimate (\candidates  and \variables\ criteria)}
\label{sec:variables}

Now our system has detected decimation should be triggered, the following question is ``which variables to decimate?''
In \cite{DBLP:journals/corr/abs-0709-1667}, the variable is chosen randomly in a uniform manner, while in \cite{Mooij_libDAI_10}, the variable with a the maximum entropy over its marginal values (the most \emph{determined} variable) is selected. Obviously, exploiting the marginal values, build throughout propagation is a good idea.

\subsubsection{From which subset choosing the candidate variables to decimate?} Both \cite{DBLP:journals/corr/abs-0709-1667} and \cite{Mooij_libDAI_10} select the only variable to decimate amongst the whole set of non-decimated variables (cf. line \ref{line:random} in Algorithm~\ref{alg:bp-guided}). Here, \candidates criterion is specified as follows:

\begin{equation}% \forall s \in \mathfrak{S}(FG),
  \label{eq:all}
  \candidates_{\mathtt{all}}(s) \eqdef \mathcal{X}\setminus\mathcal{U}
\end{equation}

However, this selection on the whole set of variables can be discussed when using local decimation triggers, like loop detection. In such case, selecting the variables to decimate within the agents in the loop, or the one which detected the loop sounds better. Another approach is to consider selecting agents depending on the past state of the system. For instance, if a variable has been decimated, good future candidates for decimation could be its direct neighbors in the FG:
\begin{equation}
    % \forall s \in \mathfrak{S}(FG), 
  \candidates_{\mathtt{neighbors}}(s) \eqdef \{x \in \mathcal{X}\setminus\mathcal{U}\ |\ neighbors(x) \cap \mathcal{U} \neq \emptyset\}
\end{equation}
with $neighbors(x_i) = \{x_j \in \mathcal{X}\ |\ j\neq i, \exists e_{ik},e_{kj}\in E\}$.

% \begin{itemize}
% \item whole set of variables
% \item all candidates (e.g. loop)
% \item neighborhood (à la DSA/MGM), which may be non-candidate variables
% \end{itemize}

\subsubsection{Which criteria to decide whether the variable decimate?} Now, we have to specify the \variables criterion used to decide which candidates decimate. In \cite{DBLP:journals/corr/abs-0709-1667}, it is fully random: it does not depends on the current state of the variables. It corresponds to make each variable roll a dice and choosing the greatest draw:

\begin{equation}
  \label{eq:max-rand}
% \forall x_i\in\mathcal{X}, s \in \mathfrak{S}(FG), 
  \variables_{\mathtt{max\_rand}}(x_i,s) \eqdef \left\{
    \begin{array}{ll}
      1, & \text{if } \forall x_j\neq x_i\in\mathcal{X}, rand(x_i) > rand(x_j)\\
      0, & \text{otherwise}
    \end{array}\right.
\end{equation}
where $rand(x)$ stands for the output of a random number generator (namely $sample$) using a uniform distribution (e.g. $U[0,1]$).

In \cite{Mooij_libDAI_10}, the variable with the maximal entropy over its marginal values is selected. This means the variable for which marginal values seems to be the most informed, in the Shannon's Information Theory sense, is chosen: 

\begin{equation}
  \label{eq:max-entropy}
% \forall x_i\in\mathcal{X}, s \in \mathfrak{S}(FG), 
  \variables_{\mathtt{max\_entropy}}(x_i,s) \eqdef \left\{
    \begin{array}{ll}
      1, & \text{if } \forall x_j\neq x_i\in\mathcal{X}, H(z_i(x_i)) > H(z_j(x_j))\\
      0, & \text{otherwise}
    \end{array}\right.
\end{equation}
with $H(z_k(x_k)) = -\sum_{d\in\mathcal{D}_k}z_k(x_k)(d)\log(z_k(x_k)(d))$.  

From this, other criteria can be derived. For instance, instead of using entropy, one can consider the maximal normalized marginal value:

\begin{equation}
  \label{eq:max-marginal}
\hspace*{-0.1cm}% \forall x_i\in\mathcal{X}, s \in \mathfrak{S}(FG), 
  \variables_{\mathtt{max\_marginal}}(x_i,s) \eqdef \left\{
    \begin{array}{ll}
      1, & \text{if } \forall x_j\neq x_i\in\mathcal{X}, \displaystyle\max_{d\in\mathcal{D}_i}(z_i(x_i)(d)) > \max_{d\in\mathcal{D}_i}(z_j(x_j)(d))\\
      0, & \text{otherwise}
    \end{array}\right.
\end{equation}

If several variables can be decimated at the same time, one may consider selecting the set of variable having an entropy or a normalized marginal value greater than a given threshold, to only decimate variable which are ``sufficiently'' determined. Hence, this approach requires setting another parameter (namely \texttt{THRESHOLD}):

\begin{equation}
  \label{eq:threshold-entropy}
  % \forall x_i\in\mathcal{X}, s \in \mathfrak{S}(FG),
  \variables_{\mathtt{threshold\_entropy}}(x_i,s) \eqdef \left\{
    \begin{array}{ll}
      1, & \text{if } H(z_i(x_i)) > \mathtt{THRESHOLD}\\
      0, & \text{otherwise}
    \end{array}\right.
\end{equation}

% Another local approach is to rely on stochastic choice, like in DSA, where a variable decimates depending on a probability (which is a parameter to provide, namely \texttt{PROB}).

% \begin{equation}
%   \label{eq:individual-rand}
%   % \forall x_i\in\mathcal{X}, s \in \mathfrak{S}(FG),
%   \variables_{\mathtt{individual\_rand}}(x_i,s) \eqdef \left\{
%     \begin{array}{ll}
%       1, & \text{if } rand(x_i) > \mathtt{PROB}\\
%       0, & \text{otherwise}
%     \end{array}\right.
% \end{equation}

% Besides, in DSA-1 for instance, only one variable in a given neighborhood could change its value, which requires some coordination :

% \begin{equation}
%   \label{eq:local-rand}
%   % \forall x_i\in\mathcal{X}, s \in \mathfrak{S}(FG),
%   \variables_{\mathtt{local\_rand}}(x_i,s) \eqdef \left\{
%     \begin{array}{ll}
%       1, & \text{if } rand(x_i) > \mathtt{PROB} \text{ and}\\& \forall x_j\in neighbors(x_i), rand(x_j) \leq \mathtt{PROB}
%       \\
%       0, & \text{otherwise}
%     \end{array}\right.
% \end{equation}

% However, such criterion does not exploit BP and resulting marginal values at all. 
Of course, many combination of the aforementioned criteria, and other criteria could be considered in our framework. We don't discuss here criteria like in DSA which does not rely on marginal values, but on stochastic decision.

\subsubsection{Which subset of variables the decision to decimate a variable depends on?} Behind this question lies the question of coordinating the variable selection. Indeed, if computing criterion \variables does not depend on the decision of other variables, the procedure is fully distributable at low communication cost, as for policies like (\ref{eq:threshold-entropy}). At the contrary, if the decision requires to be aware of the state of other variables, as for policies like (\ref{eq:max-rand}), (\ref{eq:max-entropy}) and (\ref{eq:max-marginal}), the procedure will require some system-scale coordination messages. In \cite{DBLP:journals/corr/abs-0709-1667} and \cite{Mooij_libDAI_10}, decimation only concerns all the variables, from which only one will be chosen. This requires a global coordination, or a distributed leader election protocol which may require an underlying network (ring, spanning tree, etc.), like the one used for quiescence detection, to propagate election messages \cite{lynch96}.

In some cases, the decimation decision might be at local scale, when variables will make their decision depending on the decision of their direct neighbors, or variables in the same loop. In this case, less coordination messages will be required. For instance, if considering decimating variables in a loop, only variables in the loop will implement a leader election protocol. All policies, from (\ref{eq:max-rand}) to (\ref{eq:max-marginal}), could be extended in the same manner, by replacing $\mathcal{X}$ by $loop(x_i)$, $neighours(x_i)$, or any subset of $\mathcal{X}$. For instance:

\begin{equation}
  \label{eq:max-rand-loop}
\hspace*{-0.1cm}% \forall x_i\in\mathcal{X}, s \in \mathfrak{S}(FG), 
  \variables_{\mathtt{max\_rand\_loop}}(x_i,s) \eqdef \left\{
    \begin{array}{ll}
      1, & \text{if } \forall x_j\neq x_i\in loop(x_i), rand(x_i) > rand(x_j)\\
      0, & \text{otherwise}
    \end{array}\right.
\end{equation}

% \begin{itemize}
% \item Max entropy (like \cite{DBLP:journals/corr/abs-0709-1667}) [X]
% \item Max normalized value [X]
%   % \item Max first-second difference
% \item Random Choice [X] -> DSA
% \item Max Random Sort [X] -> Montanari
% \end{itemize}

\subsection{Deciding  the Values to Assign To Decimated Variables (\values\ criterion)}
\label{sec:values}

Now variables to decimate have been selected, the question is ``which values to assign?'' Usually, in BP-based algorithms, the simplest way to select values for variables, after propagation, is to assign values with maximal marginal value (or utility). \cite{Mooij_libDAI_10} is using such a criterion for inference:

\begin{equation}
\label{eq:assign-max-marginal-deterministic}
  \values_{\mathtt{max\_marginal}}(x_i,s) \eqdef \argmax_{d\in \mathcal{D}_i} z_i(x_i)(d)
\end{equation}

While, the policy is deterministic, in \cite{DBLP:journals/corr/abs-0709-1667} the choice of the value is a random choice using the marginal values as a probability distribution:

\begin{equation}
\label{eq:assign-max-marginal-sampling}
  \values_{\mathtt{sample\_marginal}}(x_i,s) \eqdef sample(z_i(x_i))
\end{equation}

Once again, these are only some examples of policies exploiting BP, and one can easily specify many more.
% \begin{itemize}
% \item Which value to assign to decimated variables?
% \begin{itemize}
%   %\item Value with Max entropy
%   \item Value with Max normalized value [X]
%   %\item Value with Max first-second difference
%   \item Deterministic, uniform random choice or sampling (like
%     \cite{DBLP:journals/corr/abs-0709-1667}) [X]
%   \end{itemize}
% \end{itemize}

%%% Local Variables:
%%% mode: latex
%%% mode: flyspell
%%% ispell-local-dictionary: "american"
%%% TeX-master: "decimation17report"
%%% End:

\section{Experiments}
\label{sec:experiments}

In this section we evaluate the performance of different combinations of decimation policies in \decimaxsum, on a classical optimization model (Ising model), against classical Max-Sum \cite{Farinelli:2008:DCL:1402298.1402313} and its extension Max-Sum\_AD\_VP \cite{DBLP:conf/aamas/ZivanP12}, we have implemented in our own framework. %When possible, we compute the optimal solutions using the DPOP implementation found in the Frodo library~\cite{FRODO2}.

\subsection{Ising Model}

Since we are interested in evaluating our algorithms in the presence of strong dependencies among the values of variables, we evaluate them on Ising model which is a widely used benchmark in statistical physics \cite{1677515}. We use here the same settings than \cite{vinyals10dac}. Here, constraint graphs are rectangular grids where each binary variable $x_i$ is connected to its four closer neighbors (with toroidal links which connect opposite sides of the grid), and is constrained by a unary cost $r_i$. The weight of each binary constraint $r_{ij}$ is determined by first sampling a value $\kappa_{ij}$ from a uniform distribution $U[-\beta,\beta]$ and then assigning 
\[
r_{ij} (x_i,x_j) =
\begin{cases}
  \kappa_{ij} & \mbox{ if } x_i=x_j\\
  -\kappa_{ij} & \mbox{ otherwise }
\end{cases}
\]
The $\beta$ parameter controls the average strength of interactions. In our experiments we set $\beta$ to $1.6$. The weight for each unary constraint $r_i$ is determined by sampling $\kappa_i$ from a uniform distribution $U[-0.05, 0.05]$ and then assigning $r_i(0) = \kappa_i$ and $r_i(1) = -\kappa_i$.

\subsection{Results and Analysis}

In this section we analyse results of different \decimaxsum combinations to solve squared-shape Ising problems with side size varying from 10 to 20 (e.g. 100 to 400 variables). We implemented the following combinations:

\begin{itemize}
\item 11 \decimaxsum instances with different decimation policies using the following criteria:
  \begin{itemize}
  \item \emph{trigger policies} ($\Theta$ criterion):
    \begin{itemize}
    \item $\Theta_\texttt{converge}$ (from equation \ref{eq:converge}, noted \textsf{converge}),
    \item rate-based $\Theta_\texttt{frequency}$ (from equation \ref{eq:frequency}, noted
      \textsf{2-periodic}, \textsf{3-periodic}, \textsf{5-periodic},
      \textsf{10-periodic}, \textsf{20-periodic}, and
      \textsf{100-periodic}),
    \item  budget-based $\Theta_\texttt{frequency}$
      (from equation \ref{eq:frequency}, noted \textsf{periodic}),
\end{itemize}

  \item \emph{filter policy} ($\Phi$ criterion):
    \begin{itemize}
    \item the one that selects the whole set of variables as potential
      variables to decimate (i.e. $\Phi_\mathtt{all}$ from equation \ref{eq:all}),
    \end{itemize}

  \item \emph{perform policies} ($\variables$ criterion):
    \begin{itemize}
    \item $\variables_\texttt{max\_rand}$ (from equation
      \ref{eq:max-rand}, noted \textsf{random}),
    \item $\variables_\texttt{max\_entropy}$ (from equation
      \ref{eq:max-entropy}, noted \textsf{max\_entropy}),
\end{itemize}

  \item \emph{assignment policies} ($\values$ criterion):
    \begin{itemize}
    \item deterministic $\values_\mathtt{max\_marginal}$ (from
      equation \ref{eq:assign-max-marginal-deterministic}, noted
      \textsf{deterministic}),
    \item  sampled
      $\values_\mathtt{sample\_marginal}$ (from equation
      \ref{eq:assign-max-marginal-sampling}, noted \textsf{sampling}),
\end{itemize}

  \end{itemize}
\item MaxSum, as defined in \cite{Farinelli:2008:DCL:1402298.1402313},
  \item MaxSum\_AD, as defined in \cite{DBLP:conf/aamas/ZivanP12},
  \item MaxSum\_AD\_VP, as defined in  \cite{DBLP:conf/aamas/ZivanP12},
  \item Montanari-Decimation, as defined in \cite{DBLP:journals/corr/abs-0709-1667},
  \item Mooij-Decimation, as defined in \cite{Mooij_libDAI_10}.
\end{itemize}

\begin{figure}[tb]
  \centering
  \begin{subfigure}[b]{0.49\textwidth}
    \includegraphics[width=\columnwidth]{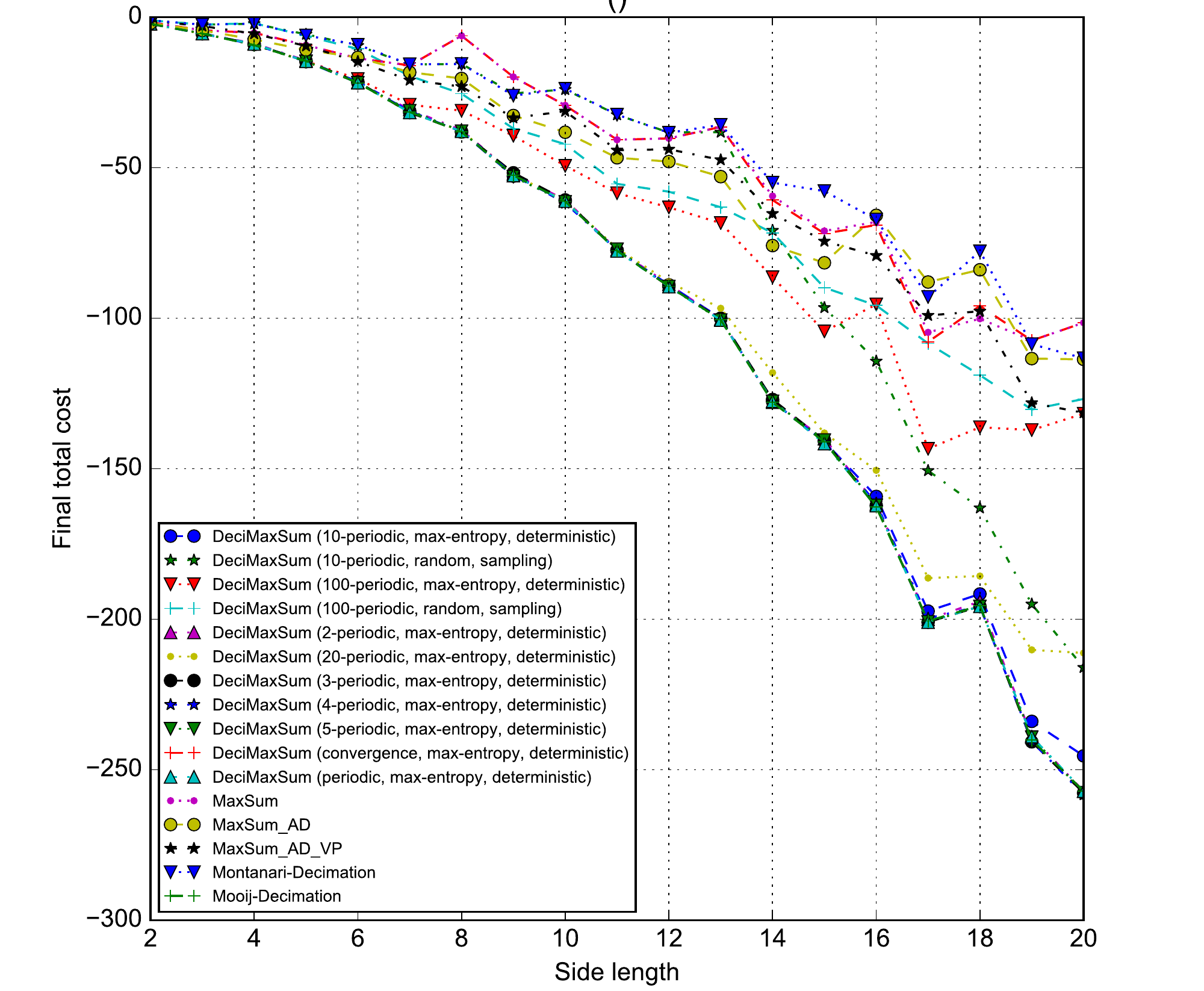}
    \caption{Final total cost}
    \label{fig:cost}  
  \end{subfigure}
% \end{figure}
% \begin{figure}[tb]
%   \centering
  \hfill
  \begin{subfigure}[b]{0.49\textwidth}
  \includegraphics[width=\columnwidth]{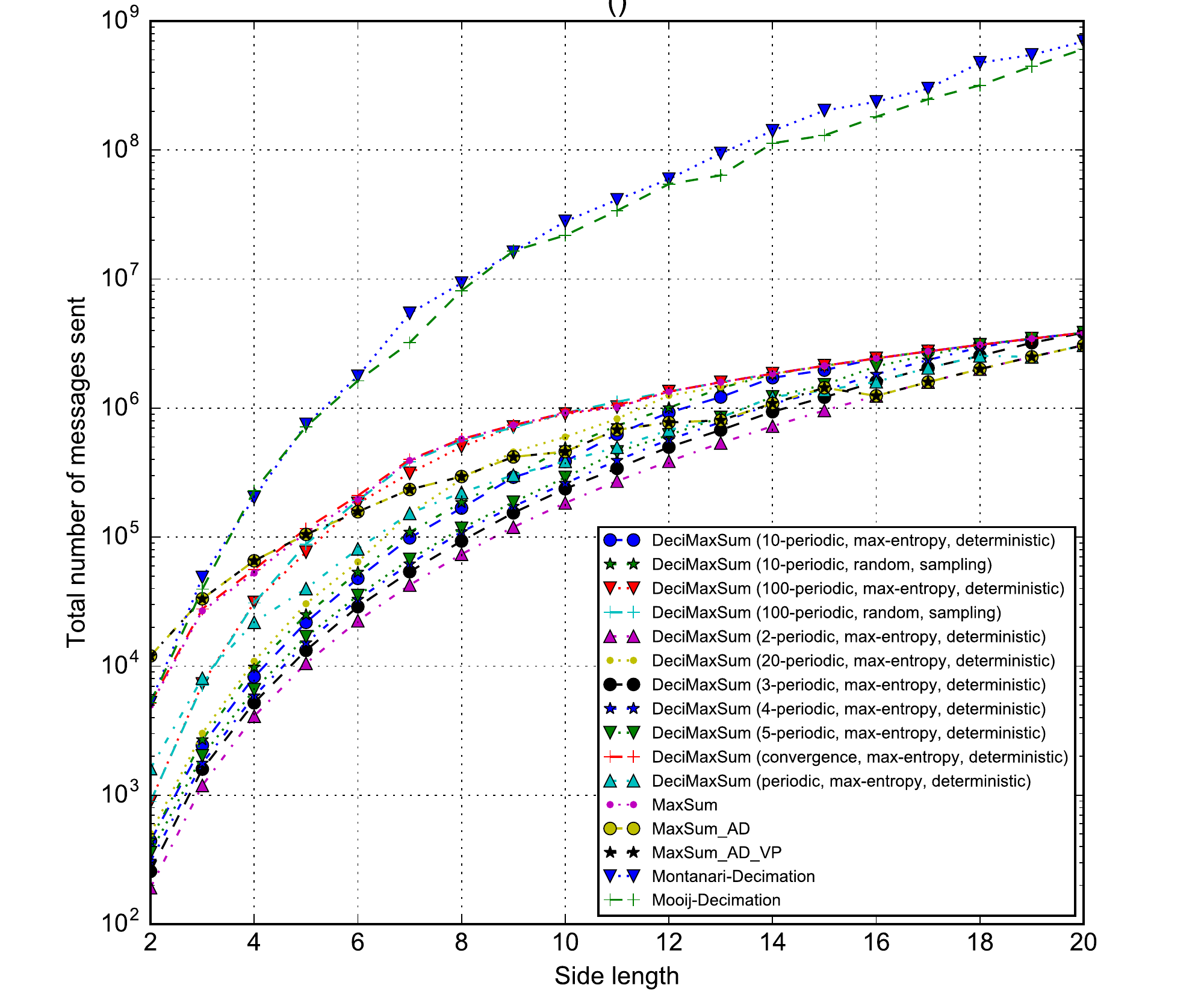}
  \caption{Final number of messages}
  \label{fig:messages}
\end{subfigure}
\caption{Performances of \decimaxsum and other solution methods on Ising problems (average of 10 problems per generator's parameter set, and average of 3 runs per problem).\label{fig:perf}}
\end{figure}

Figure~\ref{fig:perf} presents two performance metrics (final total cost and total number of exchanged messages). Considering optimality of the final solutions obtained by the different solution methods and \decimaxsum instances, what appears is that very fast decimation combined with a deterministic decimation of the most determined variable (\textsf{max\_entropy}) presents the best cost. Besides, very fast decimation also imply that few messages are exchanged compared to other solution methods, since decimation cuts message propagations. However, all the solution methods (except Montanari-Decimation and Mooij-Decimation) tend to a comparable number of exchanged messages.

%%% Local Variables:
%%% mode: latex
%%% mode: flyspell
%%% TeX-master: "decimation17report"
%%% End:

\section{Conclusions}
\label{sec:conclusion}

In this paper we have investigated how to extend Max-Sum method for solving distributed constraint optimization problems, by taking inspiration from the decimation mechanisms used to solve $k$-satisfiability problems by belief-propagation. We propose a parametric method, namely \decimaxsum, which can be set up with different decimation policies stating when to trigger decimation, which variables to decimate, and which value to assign to decimated variables. In this paper, we propose a library of such policies that can be combined to produce different versions of \decimaxsum. Our empirical results on different benchmarks show that some combinations of decimation policies outperform classical Max-Sum and its extension Max-Sum\_AD\_VP, specifically design to handle loops. \decimaxsum outputs better quality solutions in a reasonable number of message propagation. 

There are several paths to future research. First, we only explore a limited set of decimation policies. We wish to investigate more complex ones, especially policies trigger when loops are detected by agents. In fact, since our overarching goal is to cope with loops, detecting them at the agent level seems a reasonable approach to initiate decimation in a cyclic network. This approach will require agents to implement cycle-detection protocol, by sending message history, while propagating marginals. In such a setting, several decimation election may arise concurrently in the graph. Second, we would like to generalize \decimaxsum framework to consider Max-Sum\_AD\_VP as a particular case of decimation: iterated decimation. Finally, we plan to applied \decimaxsum on real world applications, with strong loopy nature, like the coordination of smart objects in IoT \cite{rust16ijcai} or decentralized energy markets in the smart grid \cite{cerquides15aamas}. 

%\acknowledgements{Work partially funded by the XXX and YYY projects.}

%%% Local Variables:
%%% mode: latex
%%% mode: flyspell
%%% TeX-master: "decimation17report"
%%% End:

\bibliographystyle{splncs03}
\bibliography{decimation17report}
\end{document}